# Low-temperature critical current of $Y_{1-x}Ca_xBa_2Cu_3O_{7-\delta}$ thin films as a function of hole content and oxygen deficiency

S. H. Naqib[a,b,c*], Anita Semwal[a]

[a]*MacDiarmid Institute for Advanced Materials and Nanotechnology, Industrial Research Ltd., P.O. Box 31310, Lower Hutt, Wellington, New Zealand*

[b]*IRC in Superconductivity, University of Cambridge, Madingley Road, Cambridge CB3 0HE, UK*

[c]*Department of Physics, Rajshahi University, Raj-6205, Rajshahi, Bangladesh*

**Abstract**

The effects of hole content (p) and oxygen deficiency (δ) on the zero-field critical current density, $J_{c0}$, were investigated for high-quality c-axis oriented $Y_{1-x}Ca_xBa_2Cu_3O_{7-\delta}$ (x = 0, 0.05, 0.10, and 0.20) thin films. Ca was used to introduce hole carriers in the $CuO_2$ planes, independent of the oxygen deficiency in the $CuO_{1-\delta}$ chains. Low-temperature $J_{c0}$(16K) of these films above the optimum doping were found to be high (> $10^7$ Amp/cm$^2$) and were primarily determined by the hole concentration, reaching a maximum at p ~ 0.185 ± 0.005, irrespective of oxygen deficiency. This implies that oxygen disorder plays only a secondary role and the intrinsic $J_{c0}$ is mainly governed by the carrier concentration and consequently by the superconducting condensation energy which also peaks at p ~ 0.19 where the pseudogap correlation vanishes.

## 1. Introduction

The importance of maximizing the superconducting critical current density, $J_c$, is one of the most important goals of applied superconductivity research. $J_c$ is limited by the depairing effect and by the flux depinning processes [1, 2]. The depairing effect is due to the breaking of the paired carriers by the induced current and depends on the superconducting condensation energy. The depinning critical current, on the other hand, is governed by the interplay between flux motion and flux pinning.

*Corresponding author. Address: IRC in Superconductivity, University of Cambridge, Madingley Road, Cambridge CB3 OHE, UK. Tel: +44-1223-337078; Fax: +44-1223-337074; E-mail: shn21@cam.ac.uk

Large critical current densities can be achieved in double $CuO_2$-layer $YBa_2Cu_3O_{7-\delta}$ (Y123) superconductor [3]. The high superconducting critical currents are clearly due to an improvement in the type and distribution of pinning centres as well as stronger c-axis coupling. Some studies have also shown that the oxygen content and hole concentration are important parameters for achieving the highest critical currents in the high temperature superconducting cuprates (HTSC) [4-6]. In a previous study it was shown that $J_c$ and the irreversibility temperature of a c-axis aligned $Y_{0.8}Ca_{0.2}Ba_2Cu_3O_{7-\delta}$ powdered sample reached a maximum in the overdoped (OD) side of the superconducting phase diagram at a hole concentration, p, of ~ 0.19 holes per planer-Cu [4]. It has been argued that p ~ 0.19 resembles a quantum critical point [4, 7] where the normal-state pseudogap correlation vanishes [7-10] and the superconducting condensate density is maximized [7, 8]. Previous studies [11, 12] also showed that oxygen content and oxygen ordering also play parts in determining the magnitude and the magnetic field and temperature dependences of $J_c$ for Y123 compounds. It is therefore of physical and technical interest to investigate the effects of hole concentration and oxygen disorder and/or deficiency on $J_c$ in Y-based double-layer cuprates. One material specific problem associated with pure Y123 compound is the fact that changes in the oxygen content and ordering always lead to a change in p, and hence the contribution to the critical current due to p or $\delta$ cannot be separated independently. This problem can be resolved by partial substitution of trivalent Y by divalent Ca. Pure Y123 with full oxygenation ($\delta$ = 0), is only slightly overdoped (p ~ 0.175, whereas superconductivity is expected to exist up to p ~ 0.27) [13-15]. Further overdoping can only be achieved by substituting $Y^{3+}$ by $Ca^{2+}$, which adds hole carriers to the $CuO_2$ planes of Y123 irrespective of the level of oxygenation of the Cu-O chains [13, 14]. Therefore, using $Y_{1-x}Ca_xBa_2Cu_3O_{7-\delta}$ compounds with different level of Ca (x) it is possible to differentiate between the contributions to the critical current due to oxygen disorder and hole concentration. In this study we report the effects of hole content and oxygen deficiency on the zero-field critical current density, $J_{c0}(T)$, for c-axis oriented $Y_{1-x}Ca_xBa_2Cu_3O_{7-\delta}$ thin films over a wide range of p, x, and δ values. We have found that the low-temperature $J_{c0}$ is primarily determined by the hole concentration, reaching a maximum at p ~ 0.185 ± 0.005, irrespective of oxygen deficiency. This indicates that the oxygen disorder and the contribution from the $CuO_{1-\delta}$ chains to the superconducting (SC) condensate (due to possibly a SC proximity effect) play a secondary role and the intrinsic $J_{c0}$ is mainly governed by the in-plane carrier concentration. $J_{c0}$ is maximized near p ~ 0.19 where the SC condensation energy also is at its largest [7, 8].

## 2. Thin film samples

High-quality c-axis oriented thin film samples of $Y_{1-x}Ca_xBa_2Cu_3O_{7-\delta}$ were synthesized from high-density single-phase sintered targets using the method of pulsed laser deposition (PLD). Thin films were grown on *(001)* $SrTiO_3$ substrates. Samples were characterized by using X-ray diffraction (XRD), atomic force microscopy (AFM), *ab*-plane room-temperature thermopower, $S_{ab}[290K]$, and *ab*-plane resistivity, $\rho_{ab}(T)$, measurements. XRD was used to determine the structural parameters, phase-purity, and degree of c-axis orientation (from the rocking-curve analysis). AFM was employed to study the grain size and the thickness of the films. All the films used in the present study were phase-pure and had high-degree of c-axis orientation (typical value of the full width at half-maximum of (007) peak was ~ $0.20^o$). Thickness of the films lies in the range (2700 ± 300) Å. $S_{ab}[290K]$ was used to calculate the planar hole content following the study by Obertelli *et al.* [16]. The level of oxygen deficiency was determined from an earlier work where the relation between δ and p were established [15]. Also, as an independent check, systematic changes in the c-axis lattice parameters were noted as δ changes as a result of oxygen annealings. $\rho_{ab}(T)$ measurements gave information regarding the impurity content and about the quality of the grain-boundaries of the films. All our samples had low values of $\rho_{ab}(300K)$ and the extrapolated zero temperature resistivity, $\rho_{ab}(0K)$ [17]. Details of the PLD method and film characterizations can be found in refs.[15, 17].

## 3. Experimental results

As prepared (AP) thin film samples were annealed at various temperatures and oxygen partial pressures to vary the oxygen content and hence the hole concentration in the $CuO_2$ planes, irrespective of the Ca content. Annealing times were kept within the range from 2 to 4 hours depending on the annealing temperatures. Samples were furnace cooled. Table-1 shows the details of the annealings and some of the outcomes. The c-axis lattice parameters after the annealings for the $Y_{0.80}Ca_{0.20}Ba_2Cu_3O_{7-\delta}$ thin film are plotted versus the oxygen deficiencies and the hole content in Figs.1a and 1b respectively. The systematic changes found here in the c-axis lattice parameter with δ are consistent with earlier studies by Jorgensen *et al.* [18] and by Hejtmánek *et al.* [19] on pure Y123 and Ca-substituted Y123 respectively. The superconducting transition temperature, $T_c$, and $J_c$ were measured using a vibrating sample magnetometer (VSM). An applied magnetic field of 0.5 mT was used for measuring $T_c$ and $J_c$ was measured for magnetic fields of up to 1 Tesla, with the applied magnetic field (H) perpendicular to the surface of the film, i.e., H||c in all cases. The zero-field critical current, $J_{c0}$, for the $Y_{1-x}Ca_xBa_2Cu_3O_{7-\delta}$ thin films with different amounts of Ca and oxygen deficiencies are shown in Figs. 2 at various temperatures. $J_{c0}$ was calculated from the magnetization loops and dimensions of the films following the method developed by Brandt and Indenbom [20].

We have shown the p-dependence of the low-T (16K) $J_{c0}$ in Fig.3a. Low-T values of $J_{c0}$ are more intrinsic in the sense that at this temperature the superconducting gap is almost fully developed and the presence of weak links have lesser contributions to the critical current. It is seen from Fig.3a that $J_{c0}$(16K) is maximized at p ~ 0.185 independent of the oxygen deficiency. $J_{c0}$(16K) versus δ plot is shown in Fig.3b, once again showing $J_{c0}$(16K) is maximized at different values of δ depending on the Ca content. Next, we have plotted the SC condensation energy, $U_0$, obtained from earlier specific heat measurements on $Y_{0.80}Ca_{0.20}Ba_2Cu_3O_{7-\delta}$ [21] and $J_{c0}$(16K) of $Y_{0.80}Ca_{0.20}Ba_2Cu_3O_{7-\delta}$ thin film versus p in Fig.4. A clear correspondence is seen between the p-dependences of $U_0$ and $J_{c0}$(16K).

## 4. Discussions and conclusion

The intrinsic depairing critical current is directly related to the superconducting condensation energy and therefore, a peak in the condensation energy should result in a peak in the critical current density, as we have indeed observed in the present study (Fig.4). Also within the collective pinning model for the magnetic flux, $J_c$ is expected to vary as ~ $U_0\xi_0$ [21], where $\xi_0$ is the zero-temperature SC coherence length. A sharp peak in $U_0(p)$ thus can enhance the flux pinning in the compound as the p-dependence of $\xi_0$ is much weaker [21]. It can be seen from Fig.3a that, qualitatively the p-dependent $J_{c0}$(16K) shows identical features independent of Ca content, namely $J_{c0}$(16K) increases with p in the underdoped side, reaches a maximum at p ~ 0.185 and then decreases with further overdoping. This implies that oxygen deficiency/disorder plays a secondary role in determining the low-T critical current density. The magnitude of the maximum $J_{c0}$(16K) for each individual film, on the other hand also show little dependence on the Ca content except for the 20%Ca substituted thin film for which the maximum $J_{c0}$(16K) is significantly lower than the other films with lesser amounts of Ca. This could be due to the lack of the contribution to the superfluid density from the Cu-O chains for the 20%Ca-Y123. Like the superconducting condensation energy, superfluid density is another measure of the strength of the superconducting pairing. Tallon *et al.* [22] found from their muon spin relaxation (μSR) study that disorder free ($\delta \rightarrow 0$) Cu-O chains enhance the superfluid density significantly. As $\delta$ increases, the chain contribution diminishes quite rapidly [22]. As it can be seen from Fig.3b, the maximum of $J_{c0}$(16K) is reached at oxygen deficiencies of ~ 0.11, ~ 0.16, and ~ 0.24 for 5%, 10%, and 20%Ca substituted thin films respectively. Therefore, the lower value of the maximum $J_{c0}$(16K) for $Y_{0.80}Ca_{0.20}Ba_2Cu_3O_{7-\delta}$ is probably due to larger chain disorder. This indicates that in order to maximize $J_c$ in Y123 large amount of Ca substitution may not be very helpful, because larger the Ca content, larger will be the oxygen deficiency required to reach p ~ 0.185. An optimal system will be the one having 5% to 10%Ca where p ~ 0.185 is attainable with relatively lower values of $\delta$.

In conclusion we have studied the p and $\delta$ dependences of the low-T zero-field critical currents of $Y_{1-x}Ca_xBa_2Cu_3O_{7-\delta}$ thin films over a wide range Ca contents. The main finding of this study is that low-T $J_{c0}$ is maximized at p ~ 0.185 irrespective of the level of oxygen content/disorder in the Cu-O chains. Low-T $J_{c0}$ is maximized at the same hole content where the superconducting condensation energy reaches its peak due to the disappearance of the normal state pseudogap correlation [7-10], which otherwise reduces both the SC condensation energy and the superfluid density by removing low-energy quasiparticle spectral weight from the SC condensate [7-10]. With further overdoping $J_{c0}$ starts decreasing following the same trend as shown by $U_0(p)$.

**Acknowledgements**

The authors would like to thank Prof. J. L. Tallon, Dr. J. R. Cooper, and Dr. G. V. M. Williams for their helpful comments and suggestions. We would also like to thank the MacDiarmid Institute for Advanced Materials and Nanotechnology, New Zealand, and the IRC in Superconductivity, University of Cambridge, UK for funding this research.

**Figure captions** (Paper title: Low-temperature critical current of $Y_{1-x}Ca_xBa_2Cu_3O_{7-\delta}$ thin films as a function of hole content and oxygen deficiency; by S. H. Naqib *et al.*)

Fig.1: The c-axis lattice parameter of $Y_{0.80}Ca_{0.20}Ba_2Cu_3O_{7-\delta}$ thin film versus (a) oxygen deficiency, $\delta$, and (b) planar hole content, p. Dashed lines are drawn as a guide to the eye.

Fig.2: Temperature dependence of the zero-field critical current density, $J_{c0}(T)$, for $Y_{1-x}Ca_xBa_2Cu_3O_{7-\delta}$ thin films for different annealings and Ca contents. (a) x = 0.0, (b) x =0.05, (c) x = 0.10, and (d) x = 0.20. Different annealing identities are shown in the plots.

Fig.3: (a) $J_{c0}$(16K) versus p for $Y_{1-x}Ca_xBa_2Cu_3O_{7-\delta}$ thin films. The dashed curves show the trends of $J_{c0}$(16K) as a function of p for the 10% and 20% Ca-substituted Y123. The vertical line shows the position of the optimum doing level where $T_c$ is maximum. (b) $J_{c0}$(16K) versus $\delta$ for $Y_{1-x}Ca_xBa_2Cu_3O_{7-\delta}$ thin films. The dashed curves show the trends of $J_{c0}$(16K) as a function of $\delta$.

Fig.4: $J_{c0}$(16K) and the superconducting condensation energy, $U_0$, versus p for $Y_{0.80}Ca_{0.20}Ba_2Cu_3O_{7-\delta}$. $U_0$ was taken from ref.[21].

**Table** (Paper title: Low-temperature critical current of $Y_{1-x}Ca_xBa_2Cu_3O_{7-\delta}$ thin films as a function of hole content and oxygen deficiency; by S. H. Naqib *et al.*)

**Table – 1: PLD conditions, annealing conditions, oxygen deficiency (δ), superconducting transition temperature ($T_c$), room-temperature thermopower (S[290K]), hole content (p), and low-temperature zero-field critical currents ($J_{c0}$(16K)) of the thin films.**

| Sample | Annealing conditions | δ (± 0.03) | $T_c$ (± 1.0)(K) | S[290K] (μV/K) | p (± 0.005) | $J_{c0}$(16K) ($10^6$ A/cm$^2$) |
|---|---|---|---|---|---|---|
| 0%Ca-Y123 | As prepared. $T_{ds}$ = 780C, $PO_2$ = 0.95 mbar. Rapidly cooled *insitu* down from $T_{ds}$ to 450C at $PO_2$ =1 atm and held for 30 mins before rapid cooling to room temp | 0.10 | 91.0 | +2.01 | 0.162 | 15.30 |
| | A1 – 575C in $O_2$ for 2 hrs, furnace cooled to room temp | 0.20 | 89.8 | +5.82 | 0.146 | 10.01 |
| | A2 – 650C in $O_2$ for 2 hrs, furnace cooled to room temp | 0.31 | 64.6 | +15.9 | 0.104 | 3.40 |
| 5%Ca-Y123 | As prepared. $T_{ds}$ = 780C, $PO_2$ = 0.95 mbar. Rapidly cooled *insitu* down from $T_{ds}$ to 450C at $PO_2$ =1 atm and held for 30 mins before rapid cooling to room temp | 0.14 | 84.4 | +0.48 | 0.170 | 16.4 |
| | B1 – 400C in $O_2$ for 3hrs, furnace cooled to room temp | 0.11 | 82.5 | -0.56 | 0.179 | 19.05 |
| | B2 – 500C in $O_2$ for 2hrs, furnace cooled to room temp | 0.18 | 85.0 | +2.59 | 0.156 | 15.8 |
| | B3 – 650C in $O_2$ for 2hrs, furnace cooled to room temp | 0.29 | 73.1 | +9.60 | 0.120 | 5.8 |
| 10%Ca-Y123 | As prepared. $T_{ds}$ = 800C, $PO_2$ = 1.20 mbar. Rapidly cooled *insitu* down from $T_{ds}$ to 450C at $PO_2$ =1 atm and held for 30 mins before rapid cooling to room temp | 0.24 | 81.4 | +2.20 | 0.160 | 15.6 |
| | C1 – 350C in $O_2$ for 4hrs, furnace cooled to room temp | 0.11 | 74.1 | -3.56 | 0.196 | 14.6 |
| | C2 – 400C in $O_2$ for 3hrs, furnace cooled to room temp | 0.16 | 78.5 | -1.40 | 0.185 | 17.4 |
| | C3 – 575C in $O_2$ for 2hrs, furnace cooled to room temp | 0.23 | 79.5 | +1.90 | 0.162 | 16.1 |
| | C4 – 650C in O2 for 2hrs, furnace cooled to room temp | 0.33 | 72.0 | +7.70 | 0.126 | 7.05 |
| 20%Ca-Y123 | As prepared. $T_{ds}$ = 800C, $PO_2$ = 1.20 mbar. Rapidly cooled *insitu* down from $T_{ds}$ at $PO_2$ =1 atm to room temp | 0.36 | 80.5 | +3.20 | 0.150 | 7.48 |
| | D1 – 350C in $O_2$ for 4hrs, furnace cooled to room temp | 0.17 | 73.1 | -3.76 | 0.197 | 11.20 |
| | D2 – 500C in $O_2$ for 2hrs, furnace cooled to room temp | 0.24 | 77.0 | -1.40 | 0.183 | 11.87 |
| | D3 – 600C in $O_2$ for 2hrs, furnace cooled to room temp | 0.30 | 81.0 | +1.23 | 0.166 | 11.60 |
| | D4 – 500C in 2% $O_2$ for 2 hrs, furnace cooled to room temp | 0.38 | 79.6 | +3.97 | 0.144 | 7.80 |
| | D5 – 550C in 2% $O_2$ for 2hrs, Furnace cooled to room temp | 0.43 | 78.0 | +5.55 | 0.136 | 6.18 |

Fig.1:

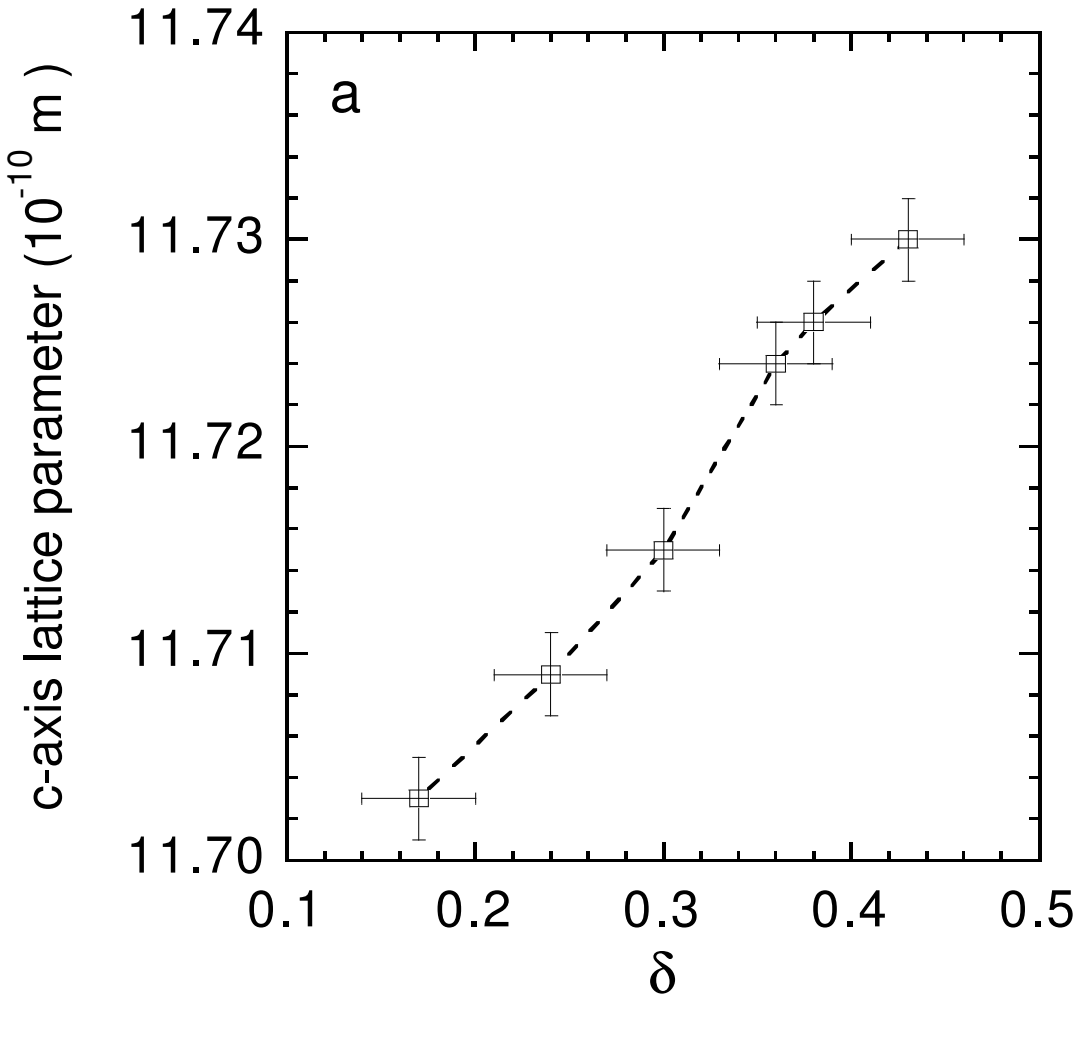

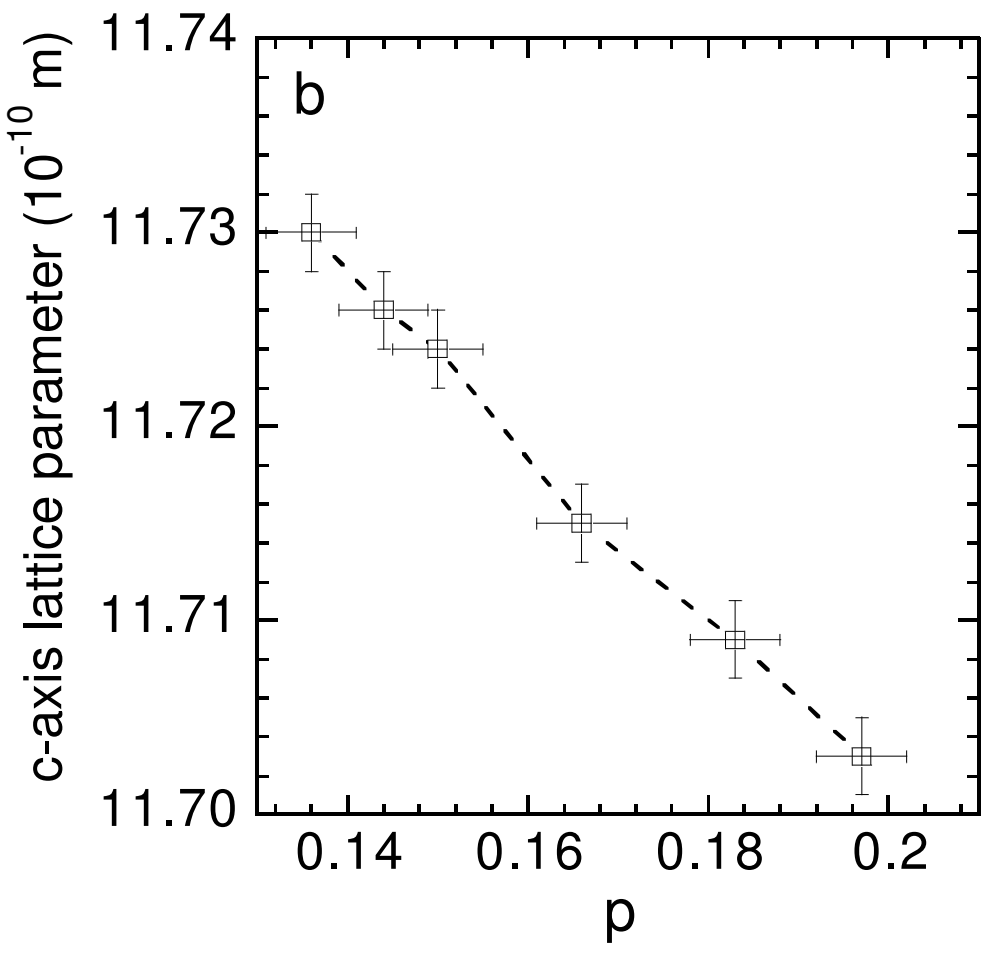

Fig.2:

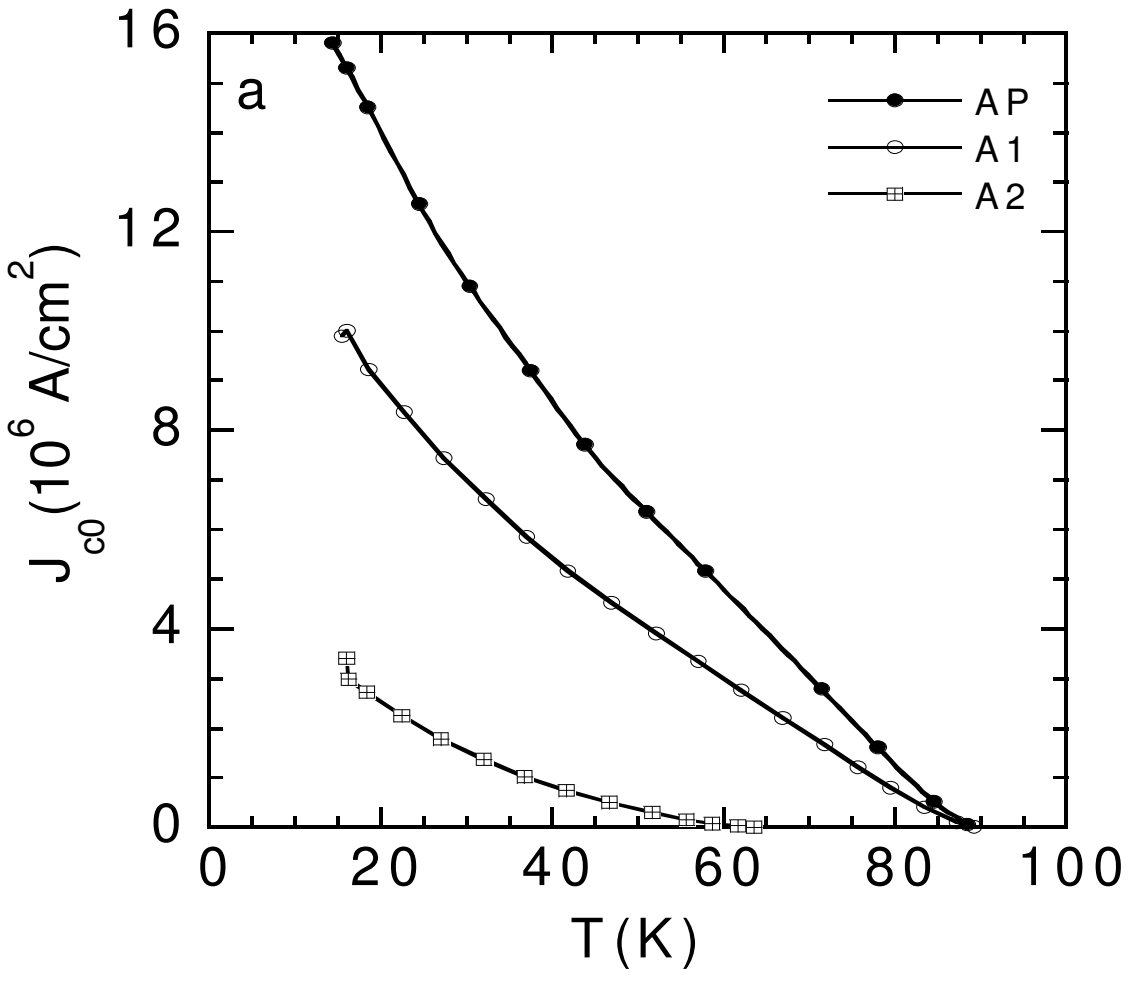

Fig.2:

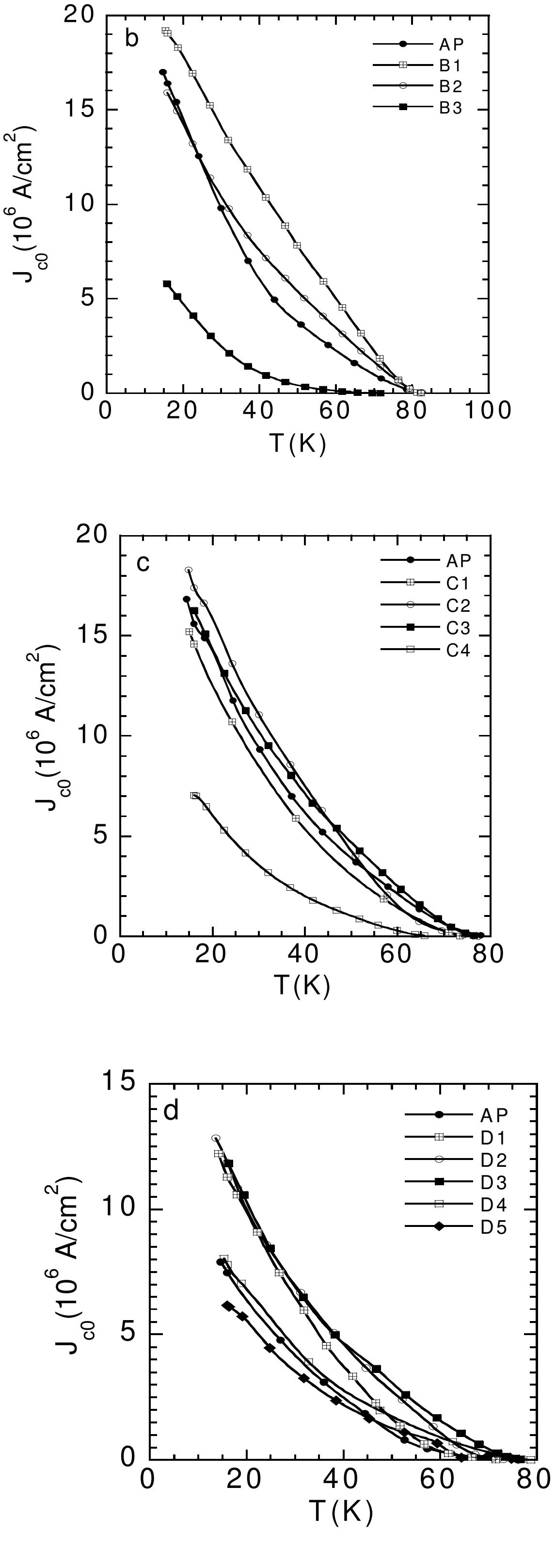

Fig.3:

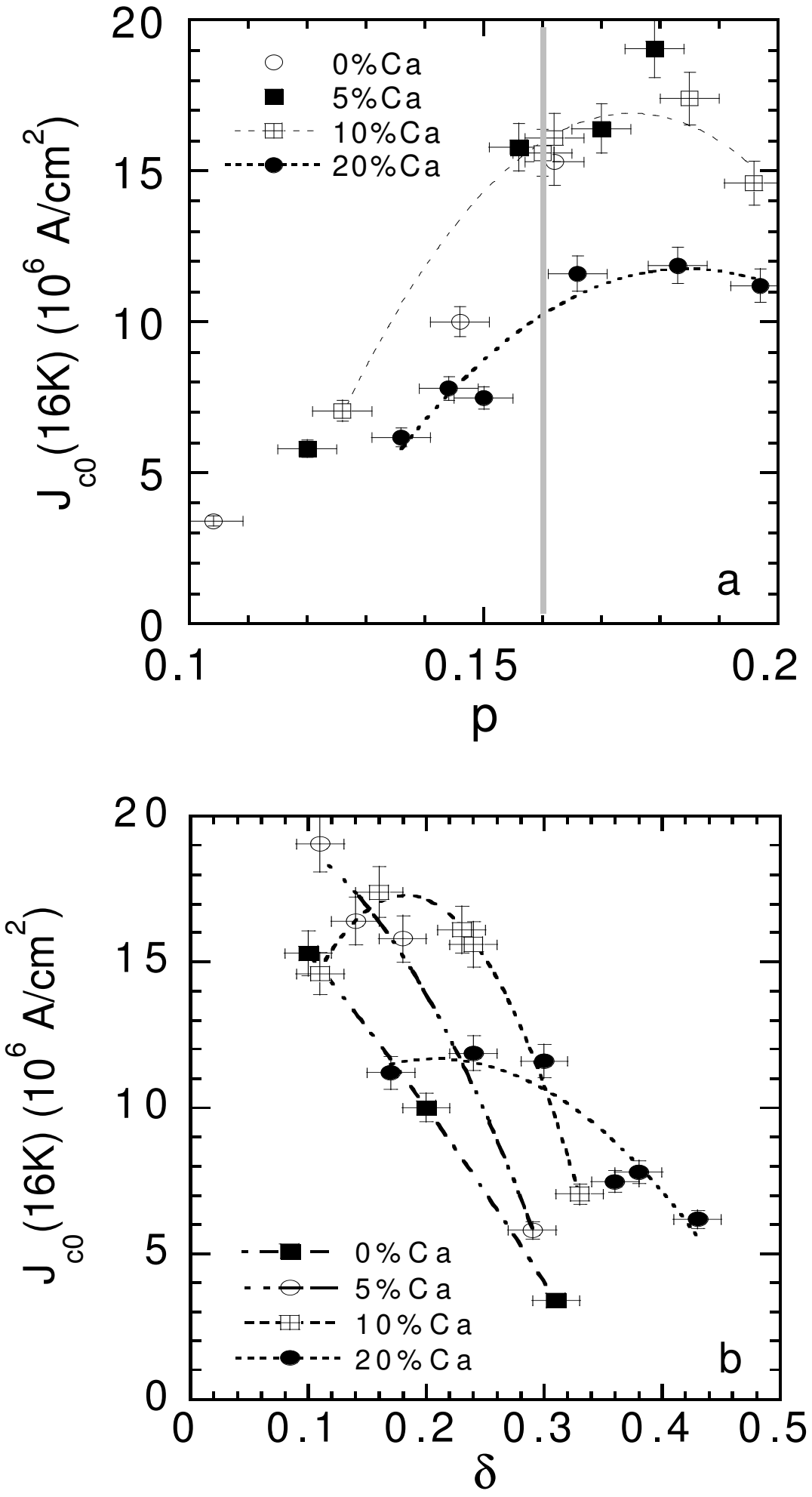

Fig.4:

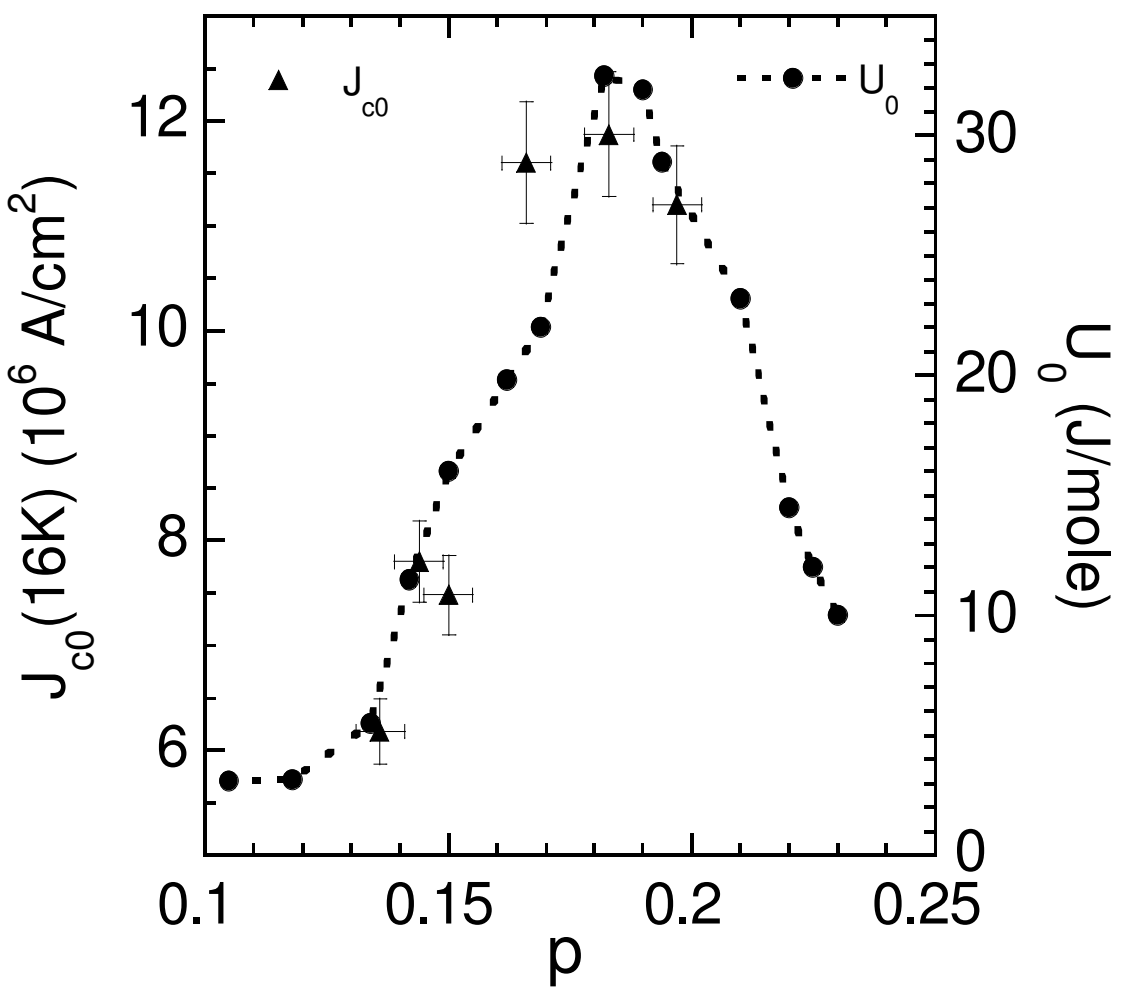

$J_{c0}$(16K) ($10^6$ A/cm$^2$)
$U_0$ (J/mole)
p
$J_{c0}$
$U_0$